\documentclass [12pt] {article}
\usepackage{xcolor}

\begin{document}

\begin{titlepage}

\title{Random-walk baryogenesis via primordial black holes}

\author{\.{I}brahim Semiz \\
{\normalsize Bo\u{g}azi\c{c}i University, Department of Physics} \\ {\normalsize Bebek 34342, \.Istanbul, Turkey} \\ \\
{\normalsize e-mail: semizibr@boun.edu.tr}
\\ \\ \\
{\normalsize Essay written for the Gravity Research Foundation}\\
{\normalsize 2016 Awards for Essays on Gravitation.} }

\date{ }

\maketitle

\begin{abstract}
Gravitation violates baryon number $B$: A star has a huge amount of it, while a black hole forming from the star has none. Consider primordial black holes before the hadronic annihiliation in the early universe, encountering and absorbing baryons and antibaryons: Each such absorption changes $B$ of the universe by one unit, up or down. But the absorption events are {\it uncorrelated and random}, hence they amount to a random walk in $B$-space, leading to the expectation of a net $|B|$ at the end.\\
\indent While the scale of this effect is most uncertain, it must exist. We explore some ramifications, including the change of net $|B|$ with expansion, connection with universe topology, and possible observational signatures. \end{abstract}


\end{titlepage}

Flip a fair coin a million times. Even though the probabilities for heads and tails are equal, you will very likely {\it not} get 500 000 heads and 500 000 tails. We all learned very early that we should expect a net difference \mbox{$|H-T|$} of about a thousand. The $H-T$ symmetry gets spontaneously broken, with unpredictable direction.

The breaking of another symmetry made our very existence possible. That is the symmetry between matter and antimatter. Einstein taught us that mass and energy can be converted into each other; but it also turned out that when matter is created from energy, an equal amount of antimatter --in fact, a mirror image of the matter-- must also be created. Conversely, you cannot simply make an amount of matter disappear and get energy for it, you also need to destroy an equal amount of antimatter, the mirror image, along with it. A good quantitative measure of "matter-ness" is baryon number, which for everyday matter is the number of protons plus neutrons; in general it is given by
\begin{equation}
B = \frac{1}{3}(n_{q}-n_{\bar{q}})
\end{equation}
where $n_{q}$ and $n_{\bar{q}}$ are the number of quarks and antiquarks, respectively. It is exactly conserved in the standard model of particle physics at the classical level, as well as to all perturbative orders. While violation is possible at the quantum level (via the chiral anomaly) or nonperturbatively, these effects are highly suppressed, at least at ordinary energies. One usually ventures beyond/outside the standard model when looking for baryon number violation; see \cite{snowmass,fornal} for recent status updates. Because of its conservation, we are not going to build a matter-conversion engine soon \cite{BH_ultimate}. 

This conservation and the symmetry, if they were exact, would make the dominance of matter in the universe a fact in need of an explanation: With the laws symmmetric, the universe should have started with zero net baryon number, and all the baryons and antibaryons in the hot primordial plasma [let us call it "hadronic plasma" to more precisely specify the epoch; after all, the universe is filled with plasma until recombination] should have annihiliated when the plasma cooled enough with the expansion of the universe, so that baryon-antibaryon  pairs could not be recreated from energy any more. Today's universe should be containing no matter, only photons. 

But, obviously, this is not what happened. Somehow, there was a tiny excess of baryons over antibaryons in the hadronic plasma, and that excess survived the annihiliation.  The question of the origin of that excess is called that of {\it baryogenesis}, and is usually addressed~\cite{textook} via the {\it Sakharov conditions} \cite{Sakharov}, i.e. (i) $B$-violation  (ii) $C$- and $CP$-violation and (iii) Interactions out of thermal equilibrium. $C$ is violated in the standard model (left-handed neutrino $\rightarrow$ left-handed antineutrino, which does not exist), and we know now CP is also violated~\cite{CP-violation,CP-nobel} ever so slightly; the matter-antimatter symmetry therefore corresponding to a slightly unfair coin.

As discussed, $B$-violation appears more naturally, if one ventures beyond the standard model. Such extensions, e.g. Grand Unified Theories (GUTs,~\cite{GUT}), motivated usually by aesthetic and naturalness considerations, are expected to be more relevant in the high-energy environment of the very early universe; the $B$-violating (supressed) effects in the standard model even more so. But let us instead consider the fourth interaction, gravitation. Even without a theory of quantum gravity, it is clear that gravitation violates baryon number: A star has a huge amount of it, while a black hole forming from the star has none, according to the so-called no-hair theorem~\cite{nohair}. Consequently, a black hole will absorb baryons and antibaryons with equal ease, regardless of itself having formed from collapse of baryonic or  antibaryonic matter. This should be contrasted with a black hole formed from matter with net positive charge, which will prefer to absorb negatively charged particles over positively charged ones. So, electric charge can still reach out to the universe from inside the black hole, whereas the baryon number of an absorbed particle is lost to the universe.

Consider then, primordial black holes (pbh) swimming in the hadronic plasma, that is, the hot thick soup of baryons and antibaryons before the hadronic annihiliation in the early universe: Each absorption of a baryon or antibaryon changes $B$ of the universe by one unit, up or down. But the absorption events are {\it uncorrelated and random}, hence they amount to a random walk in $B$-space; a flip of a $B$-coin. Even a fair coin can lead to a nonzero result now; we can conceive net baryon number being created even if $B$ was strictly conserved in all particle physics  interactions. That net number would remain behind after the annihiliation and form us eventually.

Evaporation of primordial black holes modifies the idea only slightly: If and when the Hawking temperature of the pbh becomes high enough for it to emit massive particles, the emission of each baryonic/antibaryonic particle will again constitute a unit jump in the baryon number of the universe. It can be argued that in some temperature regime, applying to either the temperature of the hadronic plasma, or to the Hawking temperature of an evaporating pbh, one has to think in terms of quarks rather than baryons, but this does not change the gist of the argument (recall hadronic jets in in particle colliders). However, the emission contribution to $|B|$ would be small with respect to the absorption contribution because most of the mass of the pbh would be radiated away as massless particles, especially if the pbh mass can reach macroscopic values by accretion. The only possible exception seems to be the case where the pbh is in a kind of dynamic equilibrium with the hadronic plasma, which seems highly improbable, given that the hadronic plasma is getting colder while the pbh is getting hotter.

Caveats apply. First of all, a little careful thought shows that if this were the only mechanism of baryogenesis, the baryon number density of the universe would decrease with time monotonically after the mechanism stops operating: Call the number of $B$-changing processes (absorptions, emissions) in a unit comoving volume of the universe $n$. For a comoving volume $V$ of a homogeneous universe then, $N = n V$; hence the produced $|B|$ is \mbox{$|B| \sim \sqrt{n} \sqrt{V}$}, therefore the produced comoving baryon number density $n_{B}$ is 
\begin{equation}
n_{B} = \frac{|B|}{V} \sim \frac{\sqrt{n}}{\sqrt{V}},
\end{equation}
in other words, the larger a comoving region of the universe one considers, the smaller the produced comoving baryon number density becomes. The operational way to "consider" a region of the universe is to wait for that region to come into causal contact, i.e. the horizon to grow to the size of that region. Hence, as time passes and the comoving horizon grows [inflation being left behind in time], smaller and smaller values of $n_{B}$ would be realized in the universe via partial neutralizations of baryon numbers which were created with different signs in regions of the universe which were previously not in causal contact.

This consideration provides both a possible observational signature, and complications in the evolution equations of the universe. The signature would be annihiliation photons with maximum energy of about 1 GeV (proton-neutron masses), the precise dependence of their number on redshift depending on the cosmological model. The complication is that we would not have $\rho_{\rm m} \propto 1/a^{3}$ any more, where $\rho_{\rm m}$ is the density of matter. The amount of matter in a comoving volume would decrease with time; it seems that this would make the dark energy problem even more acute.

In an open universe then, the comoving baryon number density would go to zero with time asymptotically. For a closed universe, however, the total comoving volume is finite, hence the comoving horizon cannot grow unboundedly, and the $n_{B}$ can reach a constant nonzero  value, if the universe does not collapse before that.

The second caveat is that we have no good idea of what $n$ might be; therefore if the magnitude of the effect could be relevant for our universe. For that, we would need the mass distribution of pbh's in the early universe as function of time or equivalent parameter, and integrate the $B$-changing processes both over time and pbh masses. The question of the pbh mass distribution is a hard one, and while efforts exist in the literature (see e.g.~\cite{pbh} for a recent review), it is hard to say that a generally accepted understanding exists.

To conclude, primordial black holes could cause or at least contribute to baryogenesis simply by swallowing more antibaryons than baryons by chance in the hadronic plasma. The effect is hard to analyze, but is inescapable. A feature distinguishing it from the more standard cosmological models is that it predicts that the comoving matter density decreases with time, making the idea falsifiable.

\end{document}